\documentclass[%
reprint,
groupedaddress,
runinaddress,
frontmatterverbose, 
showpacs,showkeys,
bibnotes,amsmath,amssymb,aps,
prl,
]{revtex4-1}

\usepackage[T2A]{fontenc}
\usepackage[cp1251]{inputenc}
\usepackage[english]{babel}
\usepackage{amsfonts,amssymb,amsmath}
\usepackage{amsmath,graphicx}
\usepackage{graphicx,pifont}

\usepackage{graphicx}
\usepackage{dcolumn}
\usepackage{bm}
\usepackage{textcomp}
\usepackage[mathlines]{lineno}

\begin{document}
%
\title{Exactly solvable model of uniaxial ferroelectrics}
\author{A.Yu.~Zakharov}\email[E-mail: ]{Anatoly.Zakharov@novsu.ru}
\author{M.I.~Bichurin}\email[E-mail: ]{Mirza.Bichurin@novsu.ru}
\author{N.V.~Evstigneeva}\email[E-mail: ]{nadya1203.89@mail.ru}
\affiliation{Novgorod State University, Veliky Novgorod, 173003, Russia}
%
%
\begin{abstract}
An exactly solvable lattice model with infinite-range potential is applied to uniaxial ferroelectrics. Asymptotically exact expression for free energy as a function of an order parameter at any temperatures is obtained. Effect of thermal expansion of lattice unit cell is taken into account.  The free energy expansion in powers of order parameter in the vicinity of critical point is presented. Corrections to Landau expansion are obtained. In particular, it is shown that summand with  external field contains a contribution of higher powers over order parameter.
\end{abstract}
%
\maketitle

\section{Introduction}
Phenomenological description of ferroelectrics in the vicinity of critical point outside the Ginzburg critical region based, as a rule, on the Landau phenomenological theory~\cite{Lan-Lif}. In this theory, the existence some ``order parameter'' $\psi$ postulated with additional assumption on free energy $F\left(\psi \right)$ expansion in powers of order parameter in the vicinity of critical point $T_c$
\begin{equation}\label{free-energy}
      F = \frac{a\left( T \right)}{2} \psi^2 + \frac{b\left( T \right)}{4} \psi^4  +   \frac{c\left( T \right)}{6} \psi^6 +  \cdots,
\end{equation}
where $a\left( T\right),\ b\left( T\right),\ldots $ are some phenomenological coefficients depending on temperature $T$.  Phase transition in this theory due to the $a\left( T \right)$ sign change in the critical point:
\begin{equation}\label{TcL}
    a\left( T\right) = \alpha \left( T - T_c\right), \quad \alpha > 0.
\end{equation}
The other coefficients $b\left( T \right), b\left( T \right)$ in sufficiently small vicinity $T_c$ assumed as some positive constants.

The main problems in this approach, beyond the Landau expansion justification, is determination of the coefficients $a\left( T \right)$, $b\left( T \right)$, $c\left( T \right)$, \ldots. Theoretical determination of these coefficients is closely connected to the justification of the Landau expansion problem and establishment connection between the system Hamiltonian and these coefficients (in some analogy with well known Mayer expansion in statistical thermodynamics of non-ideal gases). Determination of the Landau expansion coefficients using experimental data encounters difficulties due to narrowness of the applicability range of the expansion
\begin{equation}\label{T-Tc}
 Gi \lesssim \frac{ \left|T - T_c \right|}{T_c}  \ll 1,
\end{equation}
where $Gi$~ is the Ginzburg parameter (size of the critical region).

The aim of this paper is application of the exactly solvable lattice model with infinite-range interatomic potential~\cite{Par} to uniaxial ferroelectrics. For this model the exact expression for free energy as function of temperature and Hamiltonian parameters is obtained. Temperature dependence of dipole moment mean value is deduced. Exact expression for free energy as function of order parameter at arbitrary temperatures is obtained and exact Landau-like expansion in vicinity of the Curie point is derived. It is shown that exact expansion for this model contains essential corrections to the Landau expansion. Comparison between theory and experimental data is discussed.

\section{The model and partition function}
Uniaxial ferroelectrics has a distinguished direction for dipole moments orientation in the system; hence the dipole moments projections have two possible values  $\pm P$. Hamiltonian of this model suppose in following form
\begin{equation}\label{long}
  {\cal H}=-\frac{J}{2N}\sum_{j,k=1}^N P_j P_k - E\sum_{j=1}^N P_j,
\end{equation}
where is  $P_j$ a dichotomous variable with values $\pm P$, associated with $j$-th site, $-J/N P_j P_k$ is the {\em spacing independent} interactions between electrical dipoles $P_j$ and $P_k$, $E$ is an external electric field strength. Thus, in this model dipoles interaction energy does not depend on distance between these dipoles. 

Partition function of the system is 
\begin{equation}\label{part}
\begin{array}{r}
    {\displaystyle  Z\left(N, E \right) = \sum_{\left\{P_1,\cdots,P_N=\pm P\right\}}\exp\left[-\beta{\cal
  H}\right]  }\\
{\displaystyle =  \sum_{\left\{P_1,\cdots, P_N=\pm P\right\}} \exp \left[ \frac{\beta J}{2N} \left(  \sum_{j=1}^N P_j\right)^2  + \beta E\sum_{j=1}^N P_j \right],}
\end{array}
\end{equation}
where $\beta=1/T$.

Let us introduce the order parameter as mean value of a dipole moment $\left< P \right>$:
 \begin{equation}\label{<P>}
\left< P_1 \right> = \frac{1}{\beta N}\frac{ \partial \ln Z\left( N, E \right)} {\partial E}.
\end{equation}
Using well known Stratonovich-Hubbard identity\cite{Par} under the condition $ \mathrm{Re}\,\alpha>0$
\begin{equation}\label{p}
  \exp\left[\frac{\gamma^2}{2\alpha}\right]=\sqrt{\frac{\alpha}{2\pi}}
  \int\limits_{-\infty}^{+\infty}\exp\left[-\frac{\alpha}2x^2+\gamma
  x\right]dx 
\end{equation}
 and appropriate change of  variable calculation of the partition function~(\ref{part}) reduced to simple quadrature:
\begin{equation}\label{x-to-y}
     Z\left( N,E \right) = \frac{2^N}{\sqrt{2\pi}P} \int\limits_{-\infty}^{+\infty}
  \exp{\left\{- N f\left(y,E  \right)
  \right\}} dy,
\end{equation}
where 
\begin{equation}\label{SHf}
 f\left(y,E  \right) = \left[\frac{\left(y -\beta P E \right)^2}{2\beta J P^2} - \ln\cosh y  \right].
\end{equation}

Integral~(\ref{x-to-y}) converges by any values of the parameters $N, \beta, J, P, E$, but unfortunately its analytical calculation in elementary functions is impossible.

To perform analytical evaluation of the partition function~(\ref{x-to-y}) in thermodynamical limit $N \gg 1$, let us expand the function~(\ref{SHf}) in powers of $y$:
\begin{equation}\label{f(y)expan}
\begin{array}{r}
    {\displaystyle  f\left( y,E \right) =   \frac{\beta E^2}{2 J} - \frac{E}{J P}\, y +  \left(\frac{1}{\beta J P^2} -1 \right)\, \frac{y^2}{2}  }\\
{\displaystyle   + \frac{y^4}{12} - \frac{y^6}{45} + \frac{17y^8}{2520} + \cdots . }
\end{array}
\end{equation}
To calculate this integral in case $\left(1 - \beta J P^2 \right) > 0$ it is enough to keep in expansion~(\ref{f(y)expan}) the terms up to second power of $y$; otherwise, if $\left(1 - \beta J P^2 \right) \le 0$, it is necessary to keep the terms at least up to fourth power of $y$. Thus, the critical temperature $T_c$ is defined by the equality $\left(1 - \beta J P^2 \right) = 0 $:
\begin{equation}\label{Tc}
    T_c = JP^2.
\end{equation}
Let us introduce dimensionless temperature~$\tau = \frac{T}{T_c}$. Then we have
\begin{equation}\label{fySH}
  f\left( y, \mathcal{E} \right) = \frac{\tau}{2}\left(y - \frac{\mathcal{E}}{\tau}  \right)^2 - \ln\cosh y,
\end{equation}
where $\mathcal{E}$ is dimensionless external field equal to ratio of the dipole energy $EP$ in this external field $E$ to the critical temperature~$T_c$:
\begin{equation}\label{calE}
    \mathcal{E} = \frac{EP}{T_c}\,.
\end{equation}

The minimum point of function $f\left( y, \mathcal{E} \right)$ over variable $y$ obey the following equation:
\begin{equation}\label{f'y}
  \frac{\partial f\left(y,\mathcal{E} \right)}{\partial y} =  \tau \left(y - \frac{\mathcal{E}}{\tau} \right) - \tanh y = 0.
\end{equation}
Note, in critical point $\tau=1$ the second derivative $f_{yy}\left(y, \mathcal{E} \right) $ in presence of an external field is nonzero
\begin{equation}\label{d2f}
    \frac{\partial^{\,2} f\left(y,\mathcal{E} \right) }{\partial y^2} = \tau - \frac{1}{\left(\cosh y \right)^2} > 0,
\end{equation}
therefore integral~(\ref{x-to-y}) allows evaluation by means the Laplace method. As a result we have in thermodynamic limit
\begin{equation}\label{lnZN}
\begin{array}{r}
    {\displaystyle  \frac{\ln Z\left(N, \mathcal{E} \right)}{N} =  - f\left(y_0\left(\mathcal{E} \right), \mathcal{E} \right)   }\\
{\displaystyle  = - \left[  \frac{\tau}{2}\left(y_0 - \frac{\mathcal{E}}{\tau}  \right)^2 - \ln\cosh y_0  \right],   }
\end{array}
\end{equation}
where $y_0$ is solution of the equation~(\ref{f'y}). Note, the variables $y_0$ and $\mathcal{E}$ are interconnected by the equation~(\ref{f'y}).

\section{Order parameter and free energy}

Let us calculate the mean value of the dipole moment~$\left< P\right>$~(\ref{<P>}). With account of~(\ref{f'y}), we have:
\begin{equation}\label{<P1>}
\begin{array}{r}
    {\displaystyle  \left< P \right> = - \tau P\, \frac{d f\left(y_0, \mathcal{E} \right)}{d \mathcal{E}}  }\\
{\displaystyle    = \tau P \left(y_0 - \frac{\mathcal{E}}{\tau} \right) = P\tanh{y_0}\,.  }
\end{array}
\end{equation}

This formula establishes connection between order parameter~ $\left< P \right>$ and point of minimum of function~(\ref{fySH}). Then free energy per dipole has the following form
\begin{equation}\label{free-en}
    A = -T\, \frac{\ln Z_N}{N} = T \left[\frac{\tau}{2}\left(y_0 - \frac{H}{\tau}  \right)^2 - \ln\cosh y_0\right],
\end{equation}
where $y_0$ is solution of equation~(\ref{f'y}), connected with order parameter by relation~(\ref{<P1>}).

Let us express $y_0$ via order parameter $\left< P\right>$ and substitute the result into~(\ref{free-en})
\begin{equation}\label{order}
    y_0 = \frac 1 2 \ln \left( \frac{P + \left< P \right>}{P - \left< P \right>} \right).
\end{equation}
As a result, we have asymptotically exact expression for free energy~$A\left(\left< P\right> \right)$ via order parameter~$\left<P \right>$:
\begin{equation}\label{A(S)}
\begin{array}{r}
{\displaystyle      A\left( \left< P \right>, \mathcal{E} \right) = T \left\{\frac{\tau}{2}  \left[ \frac 1 2 \ln \left( \frac{P + \left< P \right>}{P - \left< P \right>} \right) - \frac{\mathcal{E}}{\tau} \right]^2 \right.   }\\
{\displaystyle  \left. + \frac 1 2 \ln \left( 1 - \left(\frac{\left< P \right>}{P} \right)^2 \right) \right\}.   }    
\end{array}
  \end{equation}

In terms of dimensionless order parameter $P_s\,\frac{\Omega\left(\tau \right)}{P\left( \tau\right)} = \left< P \right> / P$ ($-1 \le \sigma \le 1$) the free energy has the following form:
\begin{equation}\label{A(sigma)}
\begin{array}{c}
 {\displaystyle  A\left( \sigma, \mathcal{E} \right) = T \Biggl\{ \frac{\tau}{2} \left[ \frac 1 2 \ln \left( \frac{1 + \sigma}{1 - \sigma} \right)\right]^2} \\ 
{\displaystyle + \left(\frac 1 2 - \mathcal{E} \right)\ln\left(1 + \sigma  \right) + \left(\frac 1 2 + \mathcal{E} \right)\ln\left(1 - \sigma  \right) \Biggr\}}   
\end{array} 
\end{equation}
(quadratic with respect to external field term omitted because it does not depend on order parameter). Hence, for systems with Hamiltonian~(\ref{long}) the law of corresponding states takes place, i.e. equations of state of these systems in dimensionless variables $\tau, \sigma, \mathcal{E}$ have an identical form.

\section{Landau-like expansion}

Expansion of free energy~(\ref{A(sigma)}) in vicinity of the critical point has the following form:

\begin{equation}\label{A-8}
\begin{array}{r}
{\displaystyle   }\\
   {\displaystyle A\left( \sigma, \mathcal{E} \right) = T \left[  \frac{\tau - 1}{2}  \sigma^2 +  \frac{4\tau - 3}{12} \sigma^4 +  \frac{23 \tau - 15}{90} \sigma^6  + \ldots  \right]   }\\
{\displaystyle +  T \mathcal{E} \left[ \sigma + \frac{\sigma^3}{3} + \frac{\sigma^5}{5} + \frac{\sigma^7}{7} + \ldots \right].  } 
\end{array}
\end{equation}
There are two essential differences between this expansion and theory Landau. 
\begin{enumerate}
    \item In contrast to Landau expansion, all the coefficients in~(\ref{A-8}) depend on temperature. Moreover, the coefficient at even powers change their signs with temperature decreasing. This fact is not essential if $1 -\tau \ll 1$, but it prevented coefficients determination by experimental data beyond small vicinity of critical point. Hence, in general out of small vicinity of critical point it should  take into account not only the higher powers of $\sigma$ terms, but also change of the coefficients with temperature.
    \item The term with external field in~(\ref{A-8}) contains not only a linear over order parameter $\sigma$ contribution, but also a power series over $\sigma$.  
\end{enumerate}

\section{Spontaneous polarization}
\begin{figure}
\includegraphics[width=3.3in]{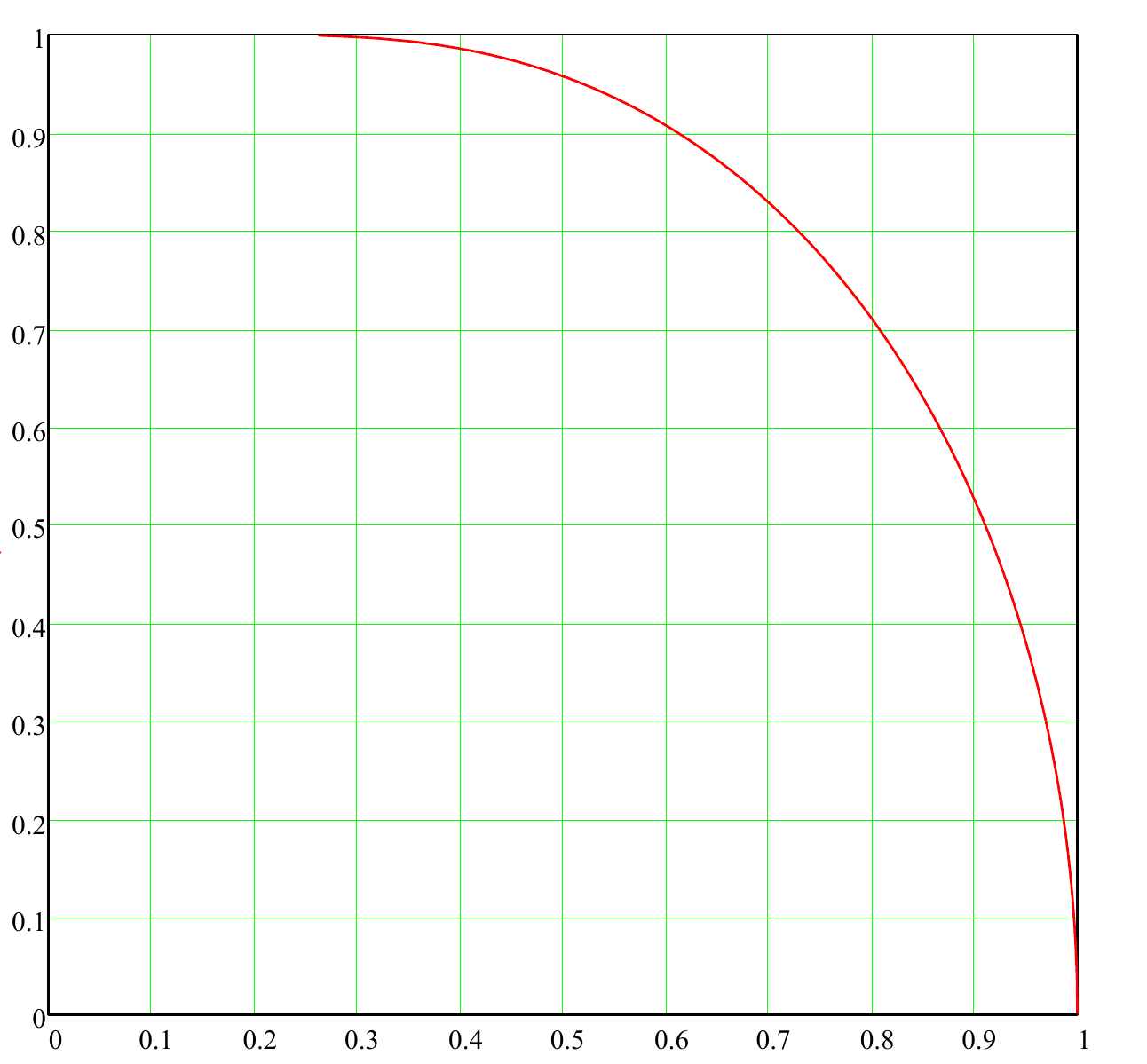} 
\caption{Dependence of dimensionless spontaneous polarization $\sigma$ on dimensionless temperature $\tau$ of uniaxial ferroelectrics with long-range potential. }
\label{fig:sigmatau}
\end{figure}

Spontaneous polarization can be found minimum free energy~(\ref{A(sigma)}) condition at~$\mathcal{E}=0$:
\begin{equation}\label{s(tau)}
    \tau = \frac{2\sigma}{\ln \left[\frac{1+\sigma}{1-\sigma} \right]}.
\end{equation}
A schematic graph of function $\sigma\left(\tau \right)$ presented on fig.~\ref{fig:sigmatau}

To perform a comparison between long-range interactions uniaxial ferroelectrics model with experimental data we should take into account some additional factors.
\begin{enumerate}
    \item We should return to dimensional polarization $P_s$ and save dimensionless temperature $\tau$.
    \item We should to take into account thermal expansion of materials and related effects. The point is that thermal expansion leads to unit cell deformation, effective ionic charges change etc.
\end{enumerate}

Note, the spontaneous polarization $P_s$ related to dipole moment mean value $\left<S \right>$ and specific volume (i.e. volume per dipole moment)  $\Omega$ via following formula
\begin{equation}\label{Ps}
    P_s = \frac{\left< P \right>}{\Omega}.
\end{equation}

Let us find the ratio $P_s\left(\tau \right)$ to $\sigma(\tau)$ with account definition $\sigma=\left< P\right>/P $:
\begin{equation}\label{Ps-sigma}
    \frac{P_s}{\sigma} = \frac{P}{\Omega} \propto \frac{q^{\star} l}{l^3} = \frac{q^{\star}}{l^2},
\end{equation}
where $l \propto \Omega^{1/3}$~is a typical linear size of dipole, $q^{\star}$~is a typical value of the effective ionic charges in dipoles. 

Change of the temperature leads to change of $l$ (thermal expansion) and $q^{\star}$ (change of the ionic contribution of chemical bonds due to redistributions of the electrons). In general, both of values $l$ and $q^{\star}$ are increasing  functions of temperature, bur there is no any certain information on the ratio $\frac{q^{\star}}{l^2}$ temperature behavior.

\begin{figure}
\includegraphics[width=3.3in]{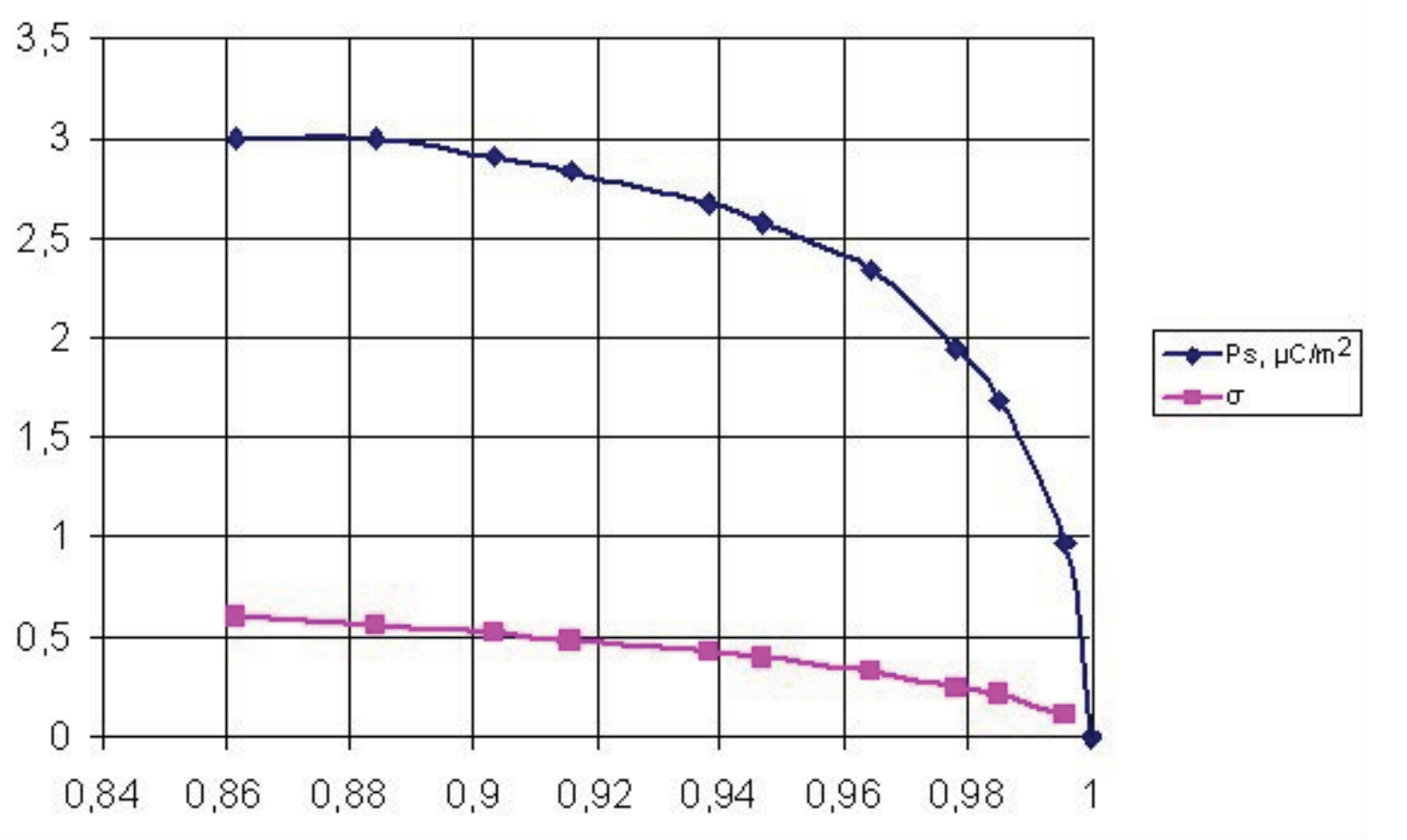} 
\caption{Dependence of triglycine-sulfate spontaneous polarization $P_s$ on dimensionless temperature $\tau$~\cite{Jona} (upper curve) and theoretical dependence of dimensionless polarization~$\sigma(\tau)$ for long-range interactions model of uniaxial ferroelectrics (lower curve).}
\label{fig:sulfate}
\end{figure}

Let us consider the experimental data on temperature dependence of spontaneous polarization $P_s(\tau)$ of the uniaxial ferroelectrics triglycine sulfate and theoretical curve for dimensionless polarization $\sigma(\tau)$ for uniaxial ferroelectrics with long-range interatomic potential. These results presented on Fig.~\ref{fig:sulfate}

\begin{figure}
\includegraphics[width=3.0in]{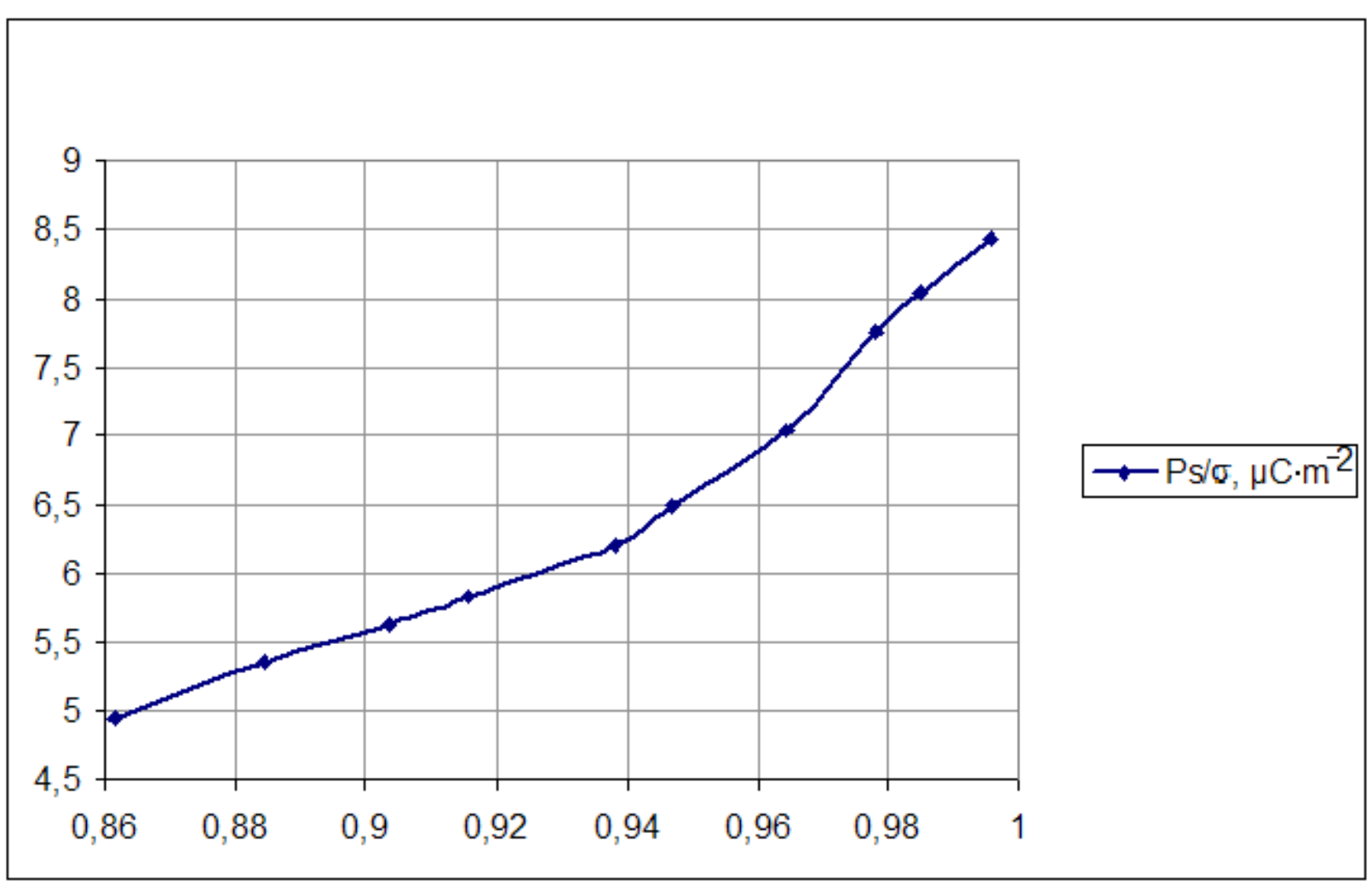} 
\caption{Dependence of the ratio $P_s\left( \tau\right)/ \sigma\left(\tau \right)$ for triglycine-sulfate.}
\label{fig:ratio}
\end{figure}

The temperature dependence of the ratio $P_s\left( \tau\right)/ \sigma\left(\tau \right)$ presented on Fig.~\ref{fig:ratio}.

The ratio $P_s\left( \tau\right)/ \sigma\left(\tau \right)$ raising with temperature increase caused, probably, by the change of ion-covalent chemical bond  to pure ionic chemical bond and related changes of the effective ionic charges in ferroelectrics. Thus, in general case variations of both the unit cell geometry and the effective ionic charges due to the materials thermal expansion should be considered.  

\section{Effects of the thermal expansion}
Let us express free energy~(\ref{A-8}) via polarization $P_s$, defined by~(\ref{Ps}):
\begin{equation}\label{s-to-Ps}
    \sigma = P_s\,\frac{\Omega\left(\tau \right)}{P\left( \tau\right)},
\end{equation} 
where $\Omega\left(\tau \right)$ and $P\left( \tau \right)$ are the unit cell volume and the dipole electric moments as functions of temperature $\tau$,  respectively. 
Substituting this relation into~(\ref{A(sigma)})  we obtain
\begin{equation}\label{A(Ps)}
\begin{array}{r}
 {\displaystyle  A\left( P_s, \mathcal{E} \right) = T \Biggl\{ \frac{\tau}{2} \left[ \frac 1 2 \ln \left( \frac{1 + P_s\,\frac{\Omega\left(\tau \right)}{P\left( \tau\right)}}{1 - P_s\,\frac{\Omega\left(\tau \right)}{P\left( \tau\right)}} \right)\right]^2} \\ 
{\displaystyle + \left(\frac 1 2 - \mathcal{E} \right)\ln\left(1 + P_s\,\frac{\Omega\left(\tau \right)}{P\left( \tau\right)}  \right) } \\ 
{\displaystyle + \left(\frac 1 2 + \mathcal{E} \right)\ln\left(1 - P_s\,\frac{\Omega\left(\tau \right)}{P\left( \tau\right)}  \right) \Biggr\}}.   
\end{array} 
\end{equation}
Spontaneous polarization $P_s$ in long-range uniaxial ferroelectrics model can be find by this expression  minimization at any values of temperature. In vicinity of the critical point $\left| 1-\tau \right| \ll 1$ the free energy expansion in powers of the polarization $P_s$ can be obtained from~(\ref{A-8}) with replacing $\sigma$ on $ \left\{\frac{\Omega\left(\tau \right)}{P\left( \tau\right)} \right\}P_s$. As a result, we obtain the following coefficients of this expansion:
\begin{equation}\label{b-c-d}
    \left\{
\begin{array}{l}
{\displaystyle a\left(T \right) = T \left(\tau -1 \right) \left\{\frac{\Omega\left(\tau \right)}{P\left( \tau\right)} \right\}^2;  }\\
{\displaystyle  b\left( T\right) =T \left( \frac{4\tau -3}{3} \right) \left\{\frac{\Omega\left(\tau \right)}{P\left( \tau\right)} \right\}^4 ; }\\
{\displaystyle c\left( T \right) = T \left( \frac{23 \tau - 15}{15} \right) \left\{\frac{\Omega\left(\tau \right)}{P\left( \tau\right)} \right\}^6 . }
\end{array}
\right.
\end{equation}
\section{Conclusion}
The exact solution of long-range uniaxial ferroelectrics model with external electric field presence is obtained. Effect of thermal expansion is taken into account. The compact expression for free energy as function of order parameter, temperature, and external electric field is derived.  It is shown that coefficients of the Landau-like expansion in critical point vicinity have essential temperature dependence, up to their signs change. It is shown that thermal expansion of unit cell leads to change of the elementary dipole moments. The last leads to additional variation of the expansion coefficients. The external field term in free energy contains not only linear with respect to order parameter summand, but some infinite series in odd powers of order parameter.

\section{Acknowledges}

We are grateful to Prof.~S.A.~Kukushkin for discussion on the paper and useful comments.
The authors gratefully acknowledge the financial support from National Science Foundation and Office of Basic Energy Science, Department of Energy, Russian Foundation for Basic Research (project No. 08-02-91359-SNF-a), Program of Russian Ministry of Education and Science ``Scientific and pedagogical personnel of innovative Russia'' on 2009.2013, Russian Ministry of Education and Science, Section 2.1.2 --- Fundamental researches in technical sciences (Project No.11324), and Department of Chemistry and Material Science of Russian Academy of Sciences.

\end{document}